\documentclass[aip,jmp,amsmath,amssymb,preprint,reprint,author-year,author-numerical]{revtex4-1}

\usepackage{graphicx}
\usepackage{dcolumn}
\usepackage{bm}

\draft 

\usepackage[hyperfigures=true,bookmarksnumbered=true,bookmarksopen=true,bookmarks=true,linkcolor=blue,urlcolor=blue,citecolor=blue,colorlinks=true,pdfhighlight=/O,pdfstartview={XYZ null null 1}]{hyperref}

\begin{document}


\title{High-resolution thermal expansion measurements under Helium-gas pressure} 



\author{Rudra Sekhar Manna}
\email[]{Manna@physik.uni-frankfurt.de}
\affiliation{Physics Institute, Goethe University Frankfurt(M), SFB/TR49, D-60438 Frankfurt(M), Germany}

\author{Bernd Wolf}
\affiliation{Physics Institute, Goethe University Frankfurt(M), SFB/TR49, D-60438 Frankfurt(M), Germany}

\author{Mariano de Souza\cite{presentaddress}}
\affiliation{Physics Institute, Goethe University Frankfurt(M), SFB/TR49, D-60438 Frankfurt(M), Germany}

\author{Michael Lang}
\affiliation{Physics Institute, Goethe University Frankfurt(M), SFB/TR49, D-60438 Frankfurt(M), Germany}


\date{\today}

\begin{abstract}

We report on the realization of a capacitive dilatometer, designed for high-resolution measurements of length changes of a material for temperatures 1.4\,K $\leq T \leq$ 300\,K and hydrostatic pressure $P \leq$ 250\,MPa. Helium ($^4$He) is used as a pressure-transmitting medium, ensuring hydrostatic-pressure conditions. Special emphasis has been given to guarantee, to a good approximation, constant-pressure conditions during temperature sweeps. The performance of the dilatometer is demonstrated by measurements of the coefficient of thermal expansion at pressures $P \simeq$ 0.1\,MPa (ambient pressure) and 104\,MPa on a single crystal of azurite, Cu$_3$(CO$_3$)$_2$(OH)$_2$, a quasi-one-dimensional spin S = 1/2 Heisenberg antiferromagnet. The results indicate a strong effect of pressure on the magnetic interactions in this system.

\end{abstract}


\maketitle 


\section{INTRODUCTION}

Dilatometry is a powerful technique for exploring the thermodynamic properties of solids. Measurements of the volume expansion coefficient, $\beta$ = ($\partial$ln$V$/$\partial T$)$_{P}$ ($V$ is the sample volume), or, when single-crystalline material is available, the uniaxial expansion coefficients, $\alpha_{i}$ = ($\partial$ln$l_{i}$/$\partial T$)$_{P}$ ($l_{i}$ denotes the sample length along the principal crystal axes) provide important information on the electronic, magnetic and lattice properties of the material. The expansivity is thermodynamically related to the specific heat via

\begin{equation}
\beta = \Gamma \cdot \kappa_{T} \cdot V_{m}^{-1}\cdot C_{V},
\label{Grueneisen}
\end{equation}

with $\kappa_{T}$ = $-$($\partial$ln$V$/$\partial P$)$_{T}$ the isothermal compressibility, $V_{m}$ the molar volume, and $\Gamma$ = $C_{V}^{-1} \cdot$ ($\partial S$/$\partial$ln$V$)$_{T}$ is the Gr\"{u}neisen function measuring the volume dependence of the system's entropy $S$. In cases where a single characteristic energy $\epsilon$ governs the properties of a material within a certain temperature range the corresponding Gr\"{u}neisen parameter is given by $\Gamma$ = $-$dln$\epsilon$/dln$V$.\cite{Barron} Hence by employing the Gr\"{u}neisen formalism the volume dependence of characteristic energies can be derived. Likewise, for a second-order phase transition at a critical temperature $T_{c}$, the application of the well-known Ehrenfest relation (dln$T_{c}$/d$P$)$_{P\rightarrow 0}$ = $V_{m}\Delta \beta/\Delta C_{P}$, enables the determination of the pressure dependence of $T_{c}$ in the limit of vanishingly small pressure from the discontinuities in $\beta$ and $C_{P}$, see \emph{e.g.} ref.\,\onlinecite{Barron} for a review.\\

Eq. \ref{Grueneisen} implies that the expansivity provides a highly sensitive probe for systems where $\Gamma$ is large or even tends to diverge. In fact large and strongly temperature-dependent $\Gamma$ values have been predicted \cite{Zhu} and experimentally observed upon approaching a quantum phase transition\cite{Kuechler, Gegenwart} or a finite-temperature critical end point.\cite{Bartosch} Hence in these cases thermal expansion measurements offer unprecedented possibilities for characterizing the phase transition. As an example of a finite-temperature critical end point where large anomalies in the expansivity are expected,\cite{deSouza, Bartosch} we mention the first-order Mott transition in organic charge-transfer salts,\cite{Kagawa} accessible by tuning the systems via pressure and/or chemical substitution. Here dilatometric measurements under well-controlled hydrostatic-pressure conditions would be of particular importance for unraveling the mystery surrounding the Mott criticality in these systems.\cite{Kagawa, Kagawa2, Papanikolaou, Bartosch} Another area of potential applications for dilatometry under He-gas pressure concerns pressure studies on materials, the properties of which show an extraordinarily strong sensitivity to nonhydrostaticity. Prominent examples include low-dimensional organic conductors \cite{Toyota} with strongly anisotropic pressure coefficients \cite{Mueller00} and some members of the recently discovered FeAs-based superconductors, notably CaFe$_{2}$As$_{2}$.\cite{Ni} For the latter material, nonhydrostatic pressure conditions were found to induce a low-temperature superconducting phase \cite{Torikachvili08} which is absent under hydrostatic pressure.\cite{Yu09} Likewise, for systems with strongly pressure-dependent magnetic exchange coupling constants, tantamount to large magnetic Gr\"{u}neisen parameters, thermal expansion measurements under pressure would provide a most sensitive tool for identifying and exploring the dominant interactions. \\

Among the various methods for measuring temperature-induced length changes of a material, capacitive dilatometry\cite{Pott} excels by its extraordinarily high sensitivity reaching values of $\Delta l/l$ = 10$^{-10}$, exceeding the resolution attainable by using strain gauges ($\Delta$\emph{l/l} $\sim$ 10$^{-6}$),\cite{Kabeya2011} X-ray diffraction ($\Delta$\emph{a/a} $\sim$ 10$^{-5}$),\cite{Barron} and optical interferometers ($\Delta$\emph{a/a} $\sim$ 10$^{-7}$-10$^{-8}$)\cite{Barron} by several orders of magnitude. In this work we describe the design and realization of an apparatus for high-resolution measurements of the coefficient of thermal expansion of solids under variable Helium-gas pressure for a wide range of temperatures 1.4\,K $\leq T \leq$ 300\,K and hydrostatic pressures $P \leq$ 250\,MPa. Since for capacitive dilatometers the pressure medium necessarily penetrates the dilatometer cell, the operation of which relies on the free motion of the sample and the capacitor plates, the experimentally accessible temperature-pressure range is restricted to the liquid (or gas) phase of $^4$He. This implies limitations of the operation range to $T \gtrsim$ 25.5\,K at $P$ = 250\,MPa and $T \gtrsim$ 14.2\,K at $P$ = 100\,MPa, for example. Apart from this constraint, liquid $^4$He provides an ideal pressure-transmitting medium as it guarantees ideal hydrostatic-pressure conditions. A first attempt for capacitive dilatometry under high gas pressure, operating under quasi-constant-pressure conditions, has been reported in ref.\,\onlinecite{Fietz}. The performance of the apparatus described here is demonstrated by dilatometric measurements both at ambient pressure and $P$ = 104\,MPa on the natural mineral azurite, Cu$_3$(CO$_3$)$_2$(OH)$_2$, a quasi-one-dimensional spin S = 1/2 Heisenberg antiferromagnet -- a good realization of a distorted diamond chain.\cite{Kikuchi} The expansivity data demonstrate a significant effect of pressure, indicating a strong pressure dependence of the magnetic interactions in this system.

\section{SYSTEM DESIGN}

\subsection{Dilatometer cell}

For the thermal expansion measurements under variable pressure conditions, a high-resolution dilatometer cell, made of high-purity (99.999\%) silver, has been used. The design of the cell is similar to the one described in ref.\,\onlinecite{Pott} except for the reduced overall dimensions of the cell, especially the diameter of the capacitor plates which are 17\,mm and 14\,mm for the lower and upper plates, respectively. For such a capacitive dilatometer, and assuming an ideal plate capacitor, the measured length change of the sample, $(\Delta\emph{l})_{meas}$, in response to a temperature change, $\Delta\emph{T}$ = $T_{2} - T_{1}$, is given by\cite{Pott}

\begin{equation}
(\Delta\emph{l})_{meas} = \epsilon_{0}\epsilon_{r}\pi\emph{r}^{2}(1+\delta)\bigg(\frac{\emph{C}_{2}-\emph{C}_{1}}{\emph{C}_{1}\cdot\emph{C}_{2}} - 2\frac{\Delta{\emph{r}}}{\emph{r}}\cdot\frac{1}{\emph{C}_{2}}\bigg),
\label{dell1}
\end{equation}

where \emph{C$_i$} = \emph{C($T_{i})$} are the capacitance values read off at the temperatures $T_{i}$, $\epsilon$$_0$ is the free space permittivity, $\epsilon_r$ is the dielectric constant of the medium between the capacitor plates of radius \emph{r}, and $\Delta r$ = $r(T_{2}) - r(T_{1}$). The constant $\delta$, accounting for inhomogeneities of the electric field at the border of the capacitor plates,\cite{Pott} can be evaluated from geometrical parameters of the dilatometer cell. A three-terminal capacitance bridge (Andeen Hagerling AH2550A) is used for measuring the capacitance \emph{C}$_i$. With the present cell, length changes as small as 5$\times$10$^{-2}$ $\AA$ can be resolved. The somewhat reduced resolution of this cell, compared to the cell described in ref. \onlinecite{Pott}, is due to the smaller size of the capacitor plates.

\subsection{Pressure cell}

\begin{figure}[!htb]
\centering
\includegraphics[width=0.50\columnwidth]{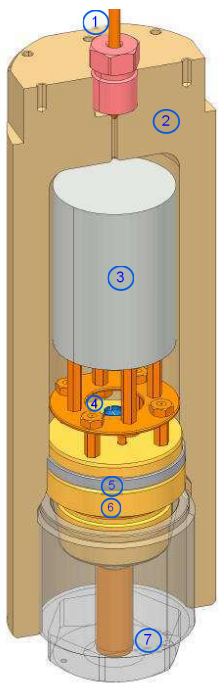}
\caption{\label{pressure-cell}(color online). Sketch of the pressure cell (made of CuBe) with an allowed maximum pressure of operation of 250\,MPa: 1 capillary with a connector on the top of the pressure cell, 2 pressure-cell body, 3 dilatometer cell, 4 n-InSb pressure sensor, 5 metal seal, 6 plug with electrical feedthroughs, 7 retaining screw. }
\end{figure}

The dilatometer cell is placed inside a pressure cell, shown in Fig.\,\ref{pressure-cell}. The cell, designed in cooperation with the Institute of High Pressure Physics, Polish Academy of Sciences, Unipress Equipment Division (abbreviated Unipress from hereon), is made of CuBe. It has an inner/outer diameter of 36/56\,mm, an inner length of 80\,mm, an inner volume of $V_{p-cell}$ = 81.4\,cm$^3$ and a weight of 2.9\,kg. The maximum pressure of operation is limited to 250\,MPa. The large volume of the cell and an inner seal diameter of 35.5\,mm is required to house not only the above-mentioned dilatometer cell of diameter 22\,mm and volume $V_{dil} \sim$ 20\,cm$^3$. It is designed for placing an even bigger cell with a diameter 33\,mm providing a higher resolution of about 10$^{-2}\,{\AA}$. For precise \textit{in-situ} measurements of the pressure, the resistivity of a doped n-InSb single crystal is measured and used as a pressure gauge, cf.\,Section $\textbf{IV.A.}$ The crystal is placed on the inner side of the plug which is unscrewed for assembling the sample into the dilatometer cell. The plug is sealed against the pressure-cell body with a brass seal coated with tin. A 60 Nm torque is needed in order to screw the retaining screw so that the seal is crushed properly between the plug and the pressure cell. The plug also includes electric feedthroughs for connecting the pressure gauge and the capacitor. The four current-type electric feedthroughs, in the form of a CuBe cone provided with a single insulated cable of 0.8 mm, are pressed into the corresponding conical holes in the plug. A single measuring-type electric feedthrough is formed by a CuBe cone, surrounded by twelve enamel-insulated Cu leads of 0.12 mm. This cone is also pressed in its corresponding conical hole. The sealing material for both cases is hygroscopic pyrophyllite. The capillary connector is sealed against the cell body using the principle cone-to-cone, widely used in high-pressure techniques.

\subsection{Cryostat insert}

\begin{figure}[!htb]
\centering
\includegraphics[width=1.0\columnwidth]{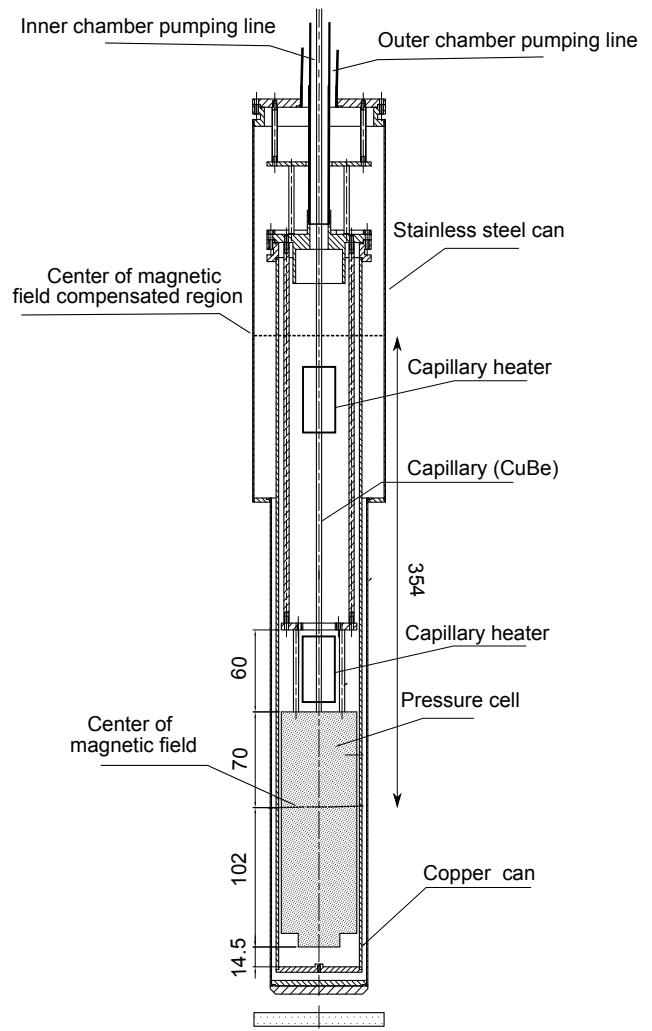}
\caption{\label{Insert} Drawing of the lower part of the home-made insert holding the pressure cell.}
\end{figure}

The pressure cell is mounted to a home-made cryostat insert, shown in Fig.\,\ref{Insert}. In designing the insert, special care has been taken to ensure smooth and well-controlled temperature sweeps of the pressure cell and the dilatometer located therein. To this end, the pressure cell is placed inside a thick-walled (2\,mm) copper can (${\O}$ 66\,mm). The can, wrapped by a thin heater foil (Kapton(R), thermofoil, Minco company) to ensure a homogeneous heat input, is surrounded by a stainless steel vacuum can. The lower tail of the vacuum can (${\O}$ 75\,mm) feeds into the bore of a superconducting solenoid providing a magnetic field of maximally 12/14\,T at 4.2/2.2\,K at the sample side. The copper and vacuum can are connected to pump lines (concentric tubes) which can be vented separately with $^4$He exchange gas. Inside the copper can the amount of exchange gas (typically 0.02\,MPa at room temperature) is kept constant throughout the whole accessible temperature range. This is sufficient to provide a good thermal contact between all parts within the can. In contrast, the amount of exchange gas inside the stainless steel can is varied, depending on the temperature range. It provides the thermal coupling of the pressure cell to the surrounding $^4$He-bath. For a given temperature interval the amount of exchange gas is adjusted such that the power applied by the temperature controller (Lakeshore LS 340) to the heater foil stays below 3.6 Watt. Capillary heaters, mounted to copper blocks, are installed around the lower part of the capillary close to the pressure cell, see Fig.\,\ref{Insert}. The heaters consist of 3 $\Omega$ resistors, connected in series, with a maximum power of 10\,Watt. In order to monitor the temperature change after switching on the capillary heater, a thermometer (Cernox resistor type CX-1080-CU) is placed on the upper copper block. For pressurizing the cell at lower temperatures, the temperature of this point of the capillary is kept about 10\,K above the temperature of the pressure cell, to avoid a blockage of the capillary due to the freezing of spurious oxygen impurities in the $^{4}$He gas. For the thermometry a pair of Cernox resistors (CX-1080-CU and CX-1050-CU), covering an overall temperature range 1.4\,K $\leq T \leq $ 300\,K, have been installed on top of the pressure cell ensuring excellent thermal anchoring to the pressure cell and the dilatometer. On demand, a second set of thermometers can be placed in the field-compensated region of the magnet, 354\,mm above the centre of the main field, where $B <$ 0.01\,T at maximum field value. These sensors enable thermometry to be performed without accounting for magnetoresistance effects of the sensors. With this setup, reproducible temperature sweeps with sweep rates of typically 1.5\,K/h can be performed, both upon decreasing and increasing temperatures. The deviations from a preset in-time linear variation are less than 0.1\,mK for $T <$ 20\,K and 0.2\,mK for 20\,K $< T <$ 300\,K.

\subsection{Pressure reservoir}

The pressure cell is connected by a CuBe capillary with an inner/outer diameter of 0.3/3\,mm and a length of $\sim$ 10\,m to an external pressure reservoir hold available at room temperature. The reservoir, required to ensure, to a good approximation, constant-pressure conditions during temperature sweeps of the pressure cell, consists of two components. (1) A standard gas bottle of volume $V_{bot}$ = 5$\times$10$^{4}$\,cm$^3$, filled with high-purity (99.999\%) $^4$He gas. This bottle is attached to (2) a compressor system, specially designed in cooperation with Unipress, to which the capillary is connected. For low pressures $P \leq$ 18\,MPa, the compressor is inactive so that the gas bottle itself serves as a reservoir. In this configuration, the large volume of the gas bottle, as compared to the pressure cell, of $V_{bot} \simeq 700\times V_{p-cell}$, ensures that the pressure inside the cell stays constant within 0.1\,MPa during temperature changes (see, \emph{e.g.}, curve (a) in Fig.\,\ref{pressure-stability} for a temperature sweep at a pressure of $P$ = 9.5\,MPa). For higher pressures $P >$ 18\,MPa, the compressor is activated and pressurizes the gas from the bottle to the desired value. The compressor consists of a low-pressure ($P \leq$ 70 MPa) and high-pressure (70\,MPa $\leq P \leq$ 250\,MPa) stage of volume $V_{LP}$ = 720\,cm$^{3}$ and $V_{HP}$ = 103\,cm$^{3}$, respectively. In addition, the compressor is equipped with an active pressure control monitoring the pressure in the high-$P$ stage. In fact, the micropump can introduce a pressure change $\Delta P/P <$ 10$^{-5}$. However, the stability of the active pressure-control system is limited by the resolution of the manometer which determines the minimum pressure change $|\Delta P|$ the system can compensate for. In the present configuration $|\Delta P|$ = 0.5\,MPa.

\subsection{Safety precautions}

As safety precautions, the gas-pressure cell and all pressure-carrying parts of the compressor are located behind safety shields in order to protect the operator. The capillary is fixed to a steel safety rope. The pressure connectors should not be tightened under pressure.

\section{CALIBRATION}

\subsection{Pressure effect on the dielectric constant of $^4$He}

In principle, eq. \ref{dell1} can also be used for analysing data taken under Helium-gas pressure. For these finite-pressure measurements, however, pressure-induced changes of the dielectric constant $\varepsilon$$_r$ of the pressure-transmitting medium $^4$He have to be taken into account. For temperatures and pressures sufficiently off the critical point of Helium at \emph{P$_c$} = 0.226\,MPa and \emph{T$_c$} = 5.2\,K,\cite{Wilks67} the variation of $\varepsilon$$_r$ with pressure can be estimated by using the Clausius-Mossotti equation

\begin{equation}
\frac{\varepsilon_r-1}{\varepsilon_r+2} = \frac{N_A\rho_m\alpha}{3M} = \frac{n\alpha}{3},
\label{Clausius-Mossotti01}
\end{equation}

where \emph{$N_A$} is Avogadro's number, \emph{$\rho$$_m$} is the mass density, \emph{M} is the molecular weight, $\alpha$ is the molecular polarizability, and \emph{n} is the number density.

\begin{figure}[!htb]
\centering
\includegraphics[width=1.0\columnwidth]{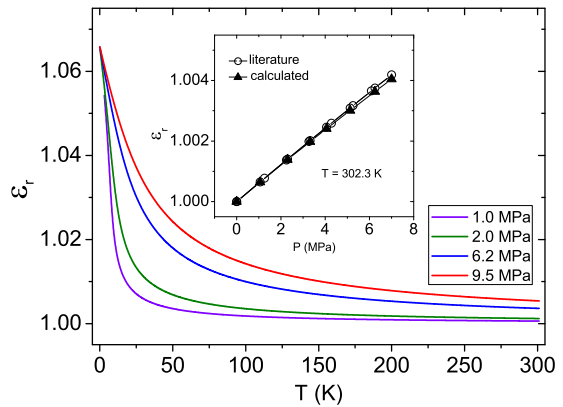}
\caption{\label{dielectric01} (color online). The variation of the dielectric constant $\varepsilon_r$ of $^4$He with temperature for various pressures calculated on the basis of eq.\,\ref{Clausius-Mossotti01} by treating $^4$He as a van der Waals gas.\cite{Manna2012} The calculations were performed by assuming a constant volume. Inset: The variation of $\varepsilon_r$ with pressure at $\emph{T}$ = 302.3\,K and fixed volume. Open circles are the literature values\cite{Schmidt} and closed triangles are the calculated values.}
\end{figure}

The Clausius-Mossotti equation works well for describing various classes of dielectric liquids and gases over wide ranges of pressure and temperatures, see, \emph{e.g.}\, refs. \onlinecite{Feynman, Rysselberghe32, Skinner72, Niemela95}. Its application to $^{4}$He, treated as a van der Waals gas with the parameters ($a$ = 0.03457\,L$^2$atm/mol$^2$ and $b$ = 0.0237\,L/mol),\cite{Weast72} for $P \leq$ 250\,MPa and 4\,K $\leq T \leq$ 300\,K yields an increase of $\varepsilon$$_r(P)$ towards lower temperatures, reaching a maximum of 1.065 at 4\,K as shown in Fig.\,\ref{dielectric01}.\cite{Manna2012} In the inset of Fig.\,\ref{dielectric01} we compare our model calculations for $\varepsilon$$_r(P)$ with experimental results, unfortunately available only in a very limited parameter range. As the figure indicates, our model reproduces well the increase of $\varepsilon$$_r$ observed for an isothermal pressure sweep at $T$ = 302.3\,K.\cite{Schmidt} At lower temperatures $T \lesssim$ 100\,K, however, test measurements on a high-purity (99.999\,\%) copper reference material, where a vanishingly small pressure effect in the expansivity is expected, see Section \textbf{III.B.} below, indicate that eq.\,\ref{Clausius-Mossotti01} is insufficient to describe $\varepsilon$$_r(P)$ with the required accuracy. Part of the inadequacy of the description applied is likely to be related to the growth of critical density fluctuations of $^{4}$He to $\varepsilon$$_r(P)$ upon approaching the critical point ($P_c, T_c$), which are not properly treated in the above Ansatz. For practical reasons, we therefore decided to account for the effect of $\varepsilon$$_r(P,T)$ by experimentally determining the ``cell effect" at varying constant-pressure conditions, which includes the effect of pressure on $\varepsilon$$_r$(\emph{P,T}).

\begin{figure}[!htb]
\centering
\includegraphics[width=1.0\columnwidth]{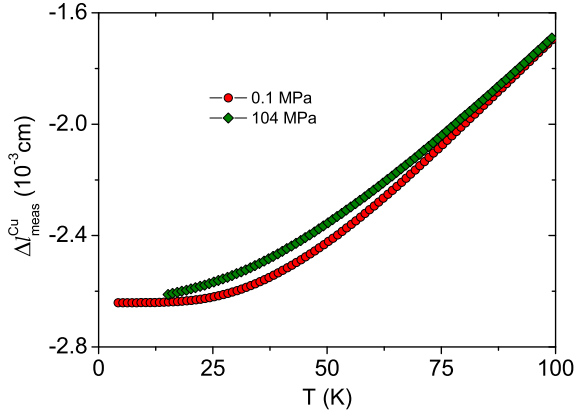}
\caption{\label{P-inducedcelleffect} (color online). Measured length changes at 0.1 MPa and 104 MPa for a copper sample used as a reference.}
\end{figure}

Fig. \ref{P-inducedcelleffect} shows length changes as a function of temperature at 0.1 MPa and 104 MPa for a copper sample used as a reference with a starting temperature $T_0$ = 300 K. The departure of the curve at 104 MPa from the data at 0.1 MPa at lower temperatures reflect the influence of pressure on  $\varepsilon$$_r$(\emph{P,T}). For the experiments on azurite, performed at the same pressure values of 0.1 MPa and 104 MPa, these data sets have been used in eqs. \ref{celleffect3} and \ref{alpha} for determining $(\Delta l/l)^{S}(T, P_{0})$ and $\alpha(T, P_0)$ (see Fig. \ref{Azurite01}).\\

\subsection{Cell effect}

In order to extract the relative length change of the sample at a given pressure from the measured data, see eq.\,\ref{celleffect3} below, all possible background effects have to be considered. This includes the thermal expansion of the dilatometer cell and all other effects resulting from the presence of the dielectric medium $^{4}$He at finite pressure. The latter includes changes of the dielectric constant as a function of pressure and temperature as well as pressure-induced changes of the geometric cell parameters. All these background effects are summarized in a generalized ``cell effect" which can be determined by measuring a proper reference material. The requirements for such a reference material are that (i) it should be a simple system with a well-known and reproducible thermal expansion, which (ii) is not affected significantly by applying pressure in the pressure range of interest. High-purity copper serves this purpose since its thermal expansivity $(\Delta l/l)^{Cu}_{lit}(T, P = 0)$ is well documented,\cite{Kroeger} and copper has a relatively small isothermal compressibility $\kappa_{T}$ at room temperature, corresponding to a bulk modulus $B_{T}$ = $\kappa_{T}^{-1}$ = 132\,GPa.\cite{Handbook69} The variation of $B_{T}$ with pressure for Cu at room temperature was determined to $(\partial B_{T}/ \partial p)_{T}$ = 5.48,\cite{Steinberg82} yielding an increase of $B_{T}$ by 0.43\% at 104\,MPa. Given that the product of the volume expansion coefficient $\beta$ and $B_{T}$ is, to a good approximation, a pressure-independent quantity for a given substance, \emph{i.e.}, $\beta(P, T)\cdot B_{T}(P, T)$ = $\beta(0, T)\cdot B_{T}(0, T)$ = $\zeta$, see ref. \onlinecite{Kumari86} and references cited therein, a correspondingly small effect on the thermal expansivity of copper at 104\,MPa can be expected. At lower temperatures an even smaller effect on $B_{T}$, and with it on $\beta$, is expected because of the hardening of the material.\\

The relative length change of the sample at a constant pressure $P_{0}$, $(\Delta l/l)^{S}(T, P_{0})$, is then given by \cite{Pott,Manna2012}:

\begin{equation}
\begin{split}
\bigg(\frac{\Delta\emph{l}}{\emph{l}}\bigg)^{S}(\emph{T,P$_{0}$}) &= \frac{1}{\emph{l}}\bigg[(\Delta\emph{l})^{S}_{meas}(\emph{T,P$_{0}$}) - (\Delta\emph{l})^{Cu}_{meas}(\emph{T,P$_{0}$})\bigg]\\
&+ \bigg(\frac{\Delta\emph{l}}{\emph{l}}\bigg)^{Cu}_{lit}(\emph{T, P = 0}),
\end{split}
\label{celleffect3}
\end{equation}

where $(\Delta l/l)^{S}_{meas}(T, P_0)$ are the measured relative length changes of the sample, which includes the cell effect, and $(\Delta l/l)^{Cu}_{meas}(T, P_0)$ are the measured relative length changes of the copper sample used as a reference.\\

The coefficient of thermal expansion $\alpha(T, P_0)$ = dln$\emph{l(T, P$_0$)}/dT$ at constant pressure $P_0$ is then approximated by

\begin{equation}
\alpha(T, P_0) \simeq \frac{[\Delta l(T_2, P_0)/l(T_2, P_0) - \Delta l(T_1, P_0)/l(T_1, P_0)]}{(T_2-T_1)}
\label{alpha}
\end{equation}

with $\Delta l(T, P_0)$ = $l(T, P_0) - l(T_0, P_0)$, $T_0$ is the starting temperature of the experiment and $T$ = $(T_1 + T_2)/2$.\\

\begin{figure}[!htb]
\centering
\includegraphics[width=1.0\columnwidth]{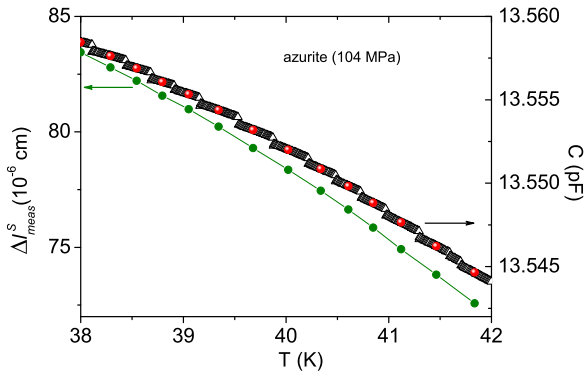}
\caption{\label{data-analysis}(color online). Variation of the measured length changes (left scale) and capacitance (right scale) upon changing the temperature in a measurement on azurite at $P$ = 104\,MPa. Step-like changes are due to the active pressure control. Red spheres represent the average of the data between two subsequent steps, with respect to both axes (\emph{T, C}).}
\end{figure}

As an example for the data analysis, we show in Fig.\,\ref{data-analysis} measured length changes of the sample (left scale) and changes of the capacitance (right scale) in a measurement performed at $P$ = 104\,MPa for the quasi-one-dimensional spin-1/2 antiferromagnet azurite, cf.\,Section \textbf{IV.B.}. A peculiarity of the data set are the step-like changes in $C$ as a function of temperature. These discontinuous changes are a consequence of the active pressure control, cf.\ Section \textbf{IV.A.}, which compensates for temperature-induced changes of the pressure once $|\Delta P|$ has reached a preset threshold of $|\Delta P|$ = 0.5\,MPa in the present case. The red spheres represent the average of the data between two subsequent pressure changes, with respect to both temperature and $C$ (or $(\Delta l)^{S}_{meas}$ obtained from eq.\,\ref{dell1}). The set of $(\Delta l/l)(T_i, P_0))$ data derived from eq.\,\ref{celleffect3} are then used to determine $\alpha (T, P_0)$ according to eq.\,\ref{alpha}.

\section{PERFORMANCE}

\subsection{Pressure determination and stability}

For a precise determination of the pressure at the sample site an n-InSb single crystal, placed inside the pressure cell, is used as a pressure gauge. The resistance of the gauge is measured in a four-terminal configuration using a Nanovoltmeter (Keithley 2182A) and a constant-current source (HP 3245A) operating at a maximum allowed current of 50\,mA. For each data point, an average is taken over typically six readouts with alternating polarity of the current in order to increase the signal-to-noise ratio and for getting rid of the effects of thermoelectric voltages. A prerequisite for its use as a pressure gauge is an accurate determination of the effect of temperature changes on the resistance at $P$ = 0, $R(P=0, T)$, of the n-InSb crystal. To this end the gauge has been calibrated as a function of temperature by using calibrated Cernox resistors (CX-1080-CU and CX-1050-CU) as a reference. At a given temperature $T$ the pressure $P(T)$, in units of GPa, is related to the resistance $R(P, T)$ according to the formula\cite{HandbookInSb}:

\begin{equation}
P(T) = [2.73 + 6.434\cdot10^{-4}\cdot(T-77.4 K)]\cdot\mathrm{ln}\bigg[\frac{R(P,T)}{R(0,T)}\bigg].
\label{empiricalformula}
\end{equation}

\begin{figure}[!htb]
\centering
\includegraphics[width=1.0\columnwidth]{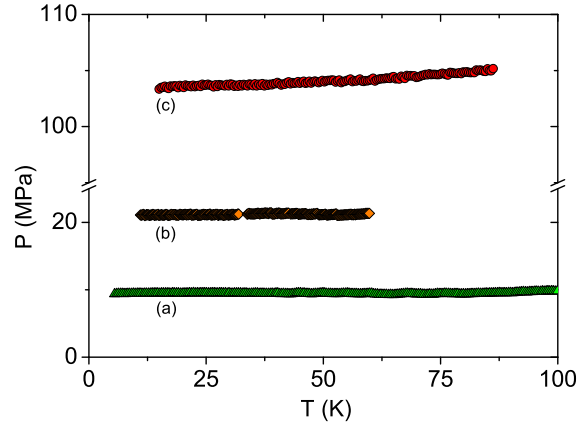}
\caption{\label{pressure-stability}(color online). Variation of pressure inside the pressure cell, as measured by the n-InSb pressure gauge, for $P$ = 9.5\,MPa (a) using the gas bottle as a reservoir, and 21\,MPa (b) as well as 104\,MPa (c) using the high-pressure compressor with active pressure control. For all measurements a sweep rate of 1.5\,K/h was used.}
\end{figure}

With this setup pressure changes within the pressure cell of $\Delta$\emph{P} $\simeq$ $\pm$ 0.1\,MPa can be resolved. In Fig.\,\ref{pressure-stability} we show the variation of pressure upon increasing the temperature for measurements performed at $P$ = 9.5\,MPa (curve (a)),  21\,MPa (curve (b)) and 104\,MPa (curve (c)). Apart from a slight increase by about 1\,MPa, corresponding to $\Delta P/P <$ 1\%, for the high-pressure experiment (curve (c)) upon increasing the temperature over a wide range from 12 to 88\,K, the data reveal an excellent stability against temperature variations.

\subsection{Thermal expansion on the distorted diamond-chain compound azurite}

\begin{figure}[!htb]
\centering
\includegraphics[width=1.0\columnwidth]{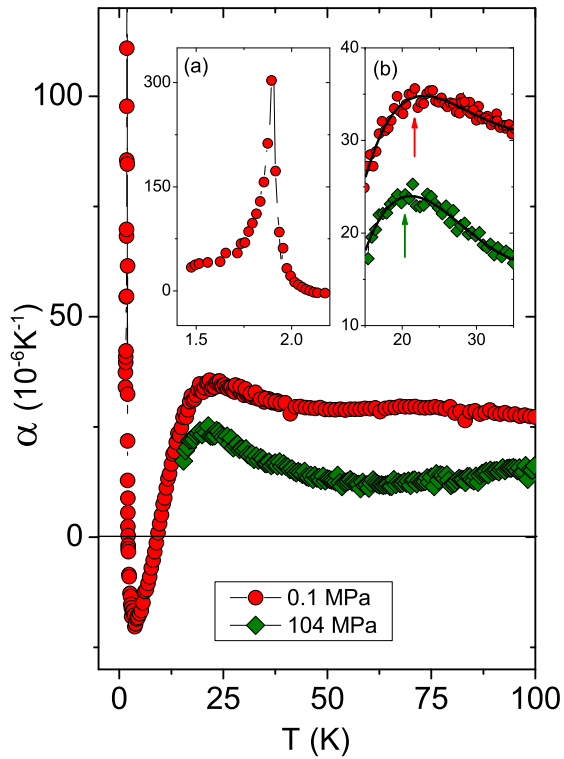}
\caption{\label{Azurite01}(color online). Uniaxial thermal expansion along the [010]-direction of azurite, Cu$_3$(CO$_3$)$_2$(OH)$_2$, measured at ambient pressure and a pressure of \emph{P} = 104\,MPa. Inset (a) shows a blow-up of the $P$ $\simeq$ 0.1 MPa data around the phase transition into long-range antiferromagnetic order. Inset (b) displays the 20\,K anomaly on expanded scales. The solid lines correspond to least squares fits to the function which serve to better visualize the shift in the position of the maximum under pressure.}
\end{figure}

The performance of the dilatometer was checked by carrying out thermal expansion measurements both at ambient pressure and $P$ = 104\,MPa Helium-gas pressure. For the latter experiment, the gas bottle in combination with the compressor was used as reservoir. The material under investigation was a single crystal of the natural mineral azurite, Cu$_3$(CO$_3$)$_2$(OH)$_2$, a quasi-one-dimensional spin S = 1/2 Heisenberg antiferromagnet, which is a good realization of a distorted diamond chain. This material has attracted considerable interest recently due to the discovery of a plateau in the low-temperature magnetization at 1/3 of the saturation magnetization.\cite{Kikuchi} The crystal studied here stems from the same large single crystal explored in ref.\,\onlinecite{Gibson}. Fig.\,\ref{Azurite01} shows the uniaxial thermal expansion coefficient along the [010]-direction, $\alpha_{b}$, of azurite for temperatures $\emph{T} \leq$ 100\,K at ambient pressure (red spheres) and at a pressure \emph{P} = 104\,MPa (olive squares). The ambient-pressure results, which coincide with published data,\cite{Andreas, Cong, Gibson} reveal three distinct anomalies, seen also in magnetic susceptibility data, which can be assigned to characteristic energy scales of azurite.\cite{Kikuchi} The rounded peak anomaly around 20\,K marks the formation of spin singlet S$_{dimer}$ = 0 dimers, involving two thirds of the material's Cu$^{2+}$ S = 1/2 spins. The corresponding antiferromagnetic intra-dimer exchange coupling $J_{2}/k_{B} \simeq$ 33\,K represents the dominant magnetic energy scale of the system.\cite{Jeschke} The minimum around 4\,K has been assigned to interacting monomers, involving the remaining other one third of the Cu$^{2+}$ spins, which form an antiferromagnetic spin-1/2 Heisenberg chain. The corresponding antiferromagnetic monomer-monomer interaction has been estimated to $J_{m}/k_{B} \simeq$ 4.6\,K.\cite{Jeschke} Finally, the phase transition anomaly at \emph{T$_N$} = 1.88\,K, shown in the inset of Fig.\,\ref{Azurite01} on expanded scales, reflects the onset of long-range antiferromagnetic order as a consequence of weak three-dimensional magnetic interactions among the monomers.\\

Fig.\,\ref{Azurite01} clearly demonstrates that an increase of the pressure from ambient pressure to \emph{P} = 104\,MPa has a strong effect on the low-temperature expansivity of azurite. Note that due to the solidification of $^4$He, the measurements at \emph{P} = 104\,MPa were limited to $\emph{T} > $ 14.5\,K.\cite{Langer} The data at \emph{P} = 104\,MPa show a surprisingly strong reduction in the absolute value of the thermal expansion coefficient compared to the ambient-pressure data. This is accompanied by a suppression of the height of the pronounced maximum in $\alpha_{b}$ around 20\,K. Since changes of the lattice expansivity under pressure of 104\,MPa are expected to be less than 1\%, cf.\,the corresponding discussion for Cu in Section $\textbf{III.B.}$, these results indicate the presence of strong magnetic contributions to $\alpha$ in the entire temperature range investigated. These magnetic contributions are most strongly suppressed at intermediate temperatures around 50\,K when pressure of 104\,MPa is applied. A thorough inspection of the data around the 20\,K maximum, cf.\,inset (b) of Fig.\,\ref{Azurite01}, discloses a slight shift of the position of the maximum to lower temperatures. In order to quantify the shift, the data have been fitted by a polynomial of fourth order. The solid lines in the inset of Fig.\,\ref{Azurite01}, corresponding to least-squares fits of these polynomials, yield a shift of the position of the maximum by about $-$1\,K at 104\,MPa. A suppression of the corresponding characteristic energy $J_{2}\cdot k_B$ under pressure is qualitatively consistent with the expectation from the Gr\"{u}neisen formalism, yielding $\Gamma_{dimer}$ = $-$dln$J_2$/dln$V$ $\propto (\beta/C_V)_{\sim 20\,K} <$ 0, since the volume expansion coefficient ($\beta$ $\simeq$ $-$13$\times$10$^{-6}$ K$^{-1}$) around this anomaly has a negative sign\cite{Andreas} while the specific heat is positive.\cite{Rule08} On a quantitative level, the present results are in excellent agreement with the observations in magnetic measurements under pressure up to about 600\,MPa \cite{Cong} yielding $\partial J_2/\partial P \simeq$ $-$1\,K/100\,MPa.

\section{Conclusions}

We have constructed and assembled a thermal expansion setup enabling high-resolution dilatometric measurements to be performed under Helium ($^4$He) gas pressure for temperatures 1.4\,K $\leq T \leq$ 300\,K and pressures $P \leq $ 250\,MPa. The system excels by a very good temperature control and a high pressure stability during temperature sweeps. The performance of the system has been demonstrated by measurements on the quasi-one-dimensional spin-1/2 Heisenberg antiferromagnet azurite, Cu$_3$(CO$_3$)$_2$(OH)$_2$, where pronounced pressure effects could be observed at low temperatures. \\

\end{document}